# Artificial Kerr effect on the self-focusing of laser in a dissipative suspension of metallic nanoparticles

N. Sepehri Javan[1,*], M. Hosseinpour Azad[1], M. N. Najafi[1]

[1]Department of Physics, University of Mohaghegh Ardabili, PO Box 179, Ardabil, Iran

[*]sepehri_javan@uma.ac.ir

**Abstract**

Self-focusing of laser beam propagating through a dissipative suspension of metallic nanoparticles is studied. Impact of imaginary part of nanoparticle polarizability on the optical force and consequently on the particles rearrangement in the presence of laser fields with an initial Gaussian profile is considered. It is shown that considering absorption of wave, leads to the creation of optical force along the wave travelling direction which can cause longitudinal distribution of nanoparticles density. Considering fifth order nonlinearity of wave amplitude that comes from simultaneous considering of normal Kerr effect produced by the inhomogeneity of refractive index resulted from ponderomotive force acting on electrons and artificial Kerr nonlinearity caused by the polarization optical force acting on particles, set of differential equations describing nonlinear steady-state evolution of laser beam is derived by using a non-paraxial method. Evolution of laser spot size for different frequencies is investigated and optimum frequency range for improving focusing property is determined. It is shown that the artificial Kerr effect causes concentration of particles near the propagation axis that can reduce threshold power for occurring self-focusing.

## I. Introduction

Interesting and exotic applications of metallic Nano-Particles (NPs) for cutting edge technologies have caused progressing interests in their synthesis and investigations of physical, chemical and mechanical properties. They have extraordinary high nonlinearity near the plasmon frequency related to the collective oscillations of relatively free electrons which makes them good candidates for nano-optics, biomedicine and nano-electronics fields where the visible light can be operated in nano-sized devices beyond the diffraction limit [1-5]. They show high-value third-order nonlinear susceptibility near the plasmon resonance that provides attractive applications in high-capacity communication networks where short response time is necessary for ultrafast switching, signal regeneration, and high speed demultiplexing [6]. Another interesting feasibility of nonlinear properties of metallic NPs is optical power limiting that can be utilized in optical sensors with high damage threshold [7] where existing of metallic NPs, especially noble metals (Au, Ag and Cu), leads to the nonlinear absorption of incident light beam in which the absorption intensively rises by the increase in the light intensity. Good experimental evidences for high nonlinear refractive index of metallic NPs near the plasmon frequency in colloidal solutions can be found in [8-11]. In addition, interaction of intense laser pulses with NPs can be the source of numerous applied phenomena. For instance, in the interaction of laser with NPs array, parametric scattering of laser can cause unstable Electromagnetic Wave (EMW) with lower frequency accompanied with the excitation of plasmon or acoustic collective oscillations via parametric Raman [12-15] or Brillouin [16] instability mechanisms. Furthermore, nonlinear interaction of laser with one-dimensional periodically nanostructured metals can result in the localization or focusing of light in an area with specific dimensions smaller than the wavelength. This phenomenon called nano-focusing. For this aim, nanostructuring of metals can be achieved by the nanosized slits including nonlinear materials [17], producing nanoholes on the surface of metallic films [18-21] or nano-grating of metallic films [22]. Recently, such a nano-focusing property is frequently investigated theoretically and experimentally in the interaction of laser with determinate number or chain of NPs [23, 24]. Moreover, localization of intense EMW can happen macroscopically during interaction of laser with a bulk medium consisting of NPs via phenomenon of Self-Focusing (SF) [25] that appears due to the change of refractive index of medium exposed to the intense EM radiation.

The study of SF in the interaction of high power lasers with nano-composites is important from the applied point of view. For example, dynamics of laser pulse in the creation of NPs by its interaction with solid targets placed in a fluid medium has a major role in the production efficiency and geometry of produced NPs as well. In this process, a longitudinally limited laser pulse focuses on a solid target placed in a liquid or gas and causes nano-sized material to be removed from the target surface. Creation of NPs in the medium can change its optical property and result in the variation of initial distribution of laser pulse in space which consequently can change the dynamics of ablation process. Even though there are some experimental works dealing with the effect of laser SF on the dynamics of laser ablation [26, 27] but to date we could not find any coherent and well-organized theoretical work about laser SF in nano-composits. Only SF of bulk medium including metallic NPs has been studied theoretically by Sepehri Javan [28] and it is recently generalized to a magnetized bulk medium of graphite nanoparticles [29]. In this paper we are studying the problem of laser SF during interaction with a suspension medium consisting of spherical metallic NPs. Using a classical microscopic theory for oscillation of NPs electronic cloud and also taking into account polarization force exerted to each neutral NP, we derive a macroscopic wave equation describing evolution of laser amplitude. Variation of the laser pulse amplitude in space is studied. An especial attention is paid on the role of artificial Kerr effect caused by the polarization force in the SF.

## II. Deriving nonlinear wave equation

Let us consider the propagation of an EMW through an aqueous medium containing metallic NPs with average radius $a$, average separation $d$, and macroscopic average density of particles $n_{c0} = 1/d^3$. We suggest that ions of each NP are immobile under interaction with the high-frequency EMW, and only the spherical electron cloud is assumed to respond to this frequency. Now, we take the electric field of the laser beam to be

$$\mathbf{E}_L \equiv \hat{E}\hat{\mathbf{x}}\cos(kz - \omega t) = \frac{1}{2}\hat{E}\hat{\mathbf{x}}e^{-i\omega t + ikz} + c.c., \qquad (1)$$

where $\hat{E}, \omega, k$ are the slowly varying amplitude, the frequency and the wave number of the laser, respectively and also the abbreviation *c.c.* denotes the complex conjugate. From Faraday's equation, the magnetic field of the laser can be obtained as

$$\mathbf{B}_L \equiv \frac{kc}{2\omega}\hat{E}\hat{\mathbf{y}}e^{-i\omega t + ikz} + c.c., \qquad (2)$$

where $c$ is the speed of light.

### II-A. Fields inside a NP

For describing interaction of laser's EMWs with particles, we calculate the microscopic fields in the place of each metallic NP taking into account the effect of polarization of particle and surrounding aqueous medium as well. For this purpose, consider a constant electric field $E_0$ aligned with the $z$ direction. The permittivity of metallic NP and ambient medium are $\varepsilon_n$ and $\varepsilon_h$, respectively. Solving Laplace's equation with boundary conditions leads to the following equations

$$\phi_n = -\left(\frac{3}{2+\varepsilon_n/\varepsilon_h}\right)E_0 r\cos\theta, \tag{3}$$

$$\phi_h = -E_0 r\cos\theta + \left(\frac{-1+\varepsilon_n/\varepsilon_h}{2+\varepsilon_n/\varepsilon_h}\right)\frac{a^3}{r^2}E_0\cos\theta, \tag{4}$$

for the electrostatic potentials inside and outside the metallic NP, respectively, where $r$ and $\theta$ are the radial coordinate and azimuthal angle of spherical coordinates system, respectively. Taking gradient of potential given by Eq. (3), leads to the electric field inside the NP as following

$$\mathbf{E_i} = \left(\frac{3}{2+\varepsilon_n/\varepsilon_h}\right)\mathbf{E_0}. \tag{5}$$

Directly from equation (5), we can find that the microscopic electric field attenuates or amplifies inside the metallic NP by the factor of

$$\alpha = \frac{3}{2+\varepsilon_n/\varepsilon_h}. \tag{6}$$

In addition, from equation (4), we can find out that the first term of the potential of outside space is the primary electric field and the second term is the contribution of particle polarization whose characteristic can be expressed by the following polarization vector

$$\mathbf{P_i} = \frac{3}{4\pi}\left(\frac{-1+\varepsilon_n/\varepsilon_h}{2+\varepsilon_n/\varepsilon_h}\right)\mathbf{E_0}. \tag{7}$$

**II-B. Particles density distribution under interaction with laser**

In contrast to lattice form of NPs, in a colloidal suspension, particles can move under interaction with fields of laser beam with finite longitudinal or transvers structure which in turn it can change initial form of laser and affects future dynamics of system. Therefore one should self-consistently solve Maxwell's equations with motion equation of both neutral NPs and conduction electrons of each NP. First we calculate density of NPs in a steady state regime in which optical gradient force (or dipole force) caused by non-homogeneity of laser intensity and diffusive force created by mutual collisions balance together. As review of related literature shows [30-32], in the most of laser-nanosuspension interaction problems, in calculating optical force, it is assumed that there is no imaginary part for permittivity of medium and it is not dissipative. Here, we consider a dissipative medium and modify the optical force acting on. Fortunately, in the case of Rayleigh regime where particle size is small compared to the wavelength, such a modification has been accomplished in Ref. [33] and the averaged optical force acting on a dissipative particle with complex polarizability of $\chi = \chi' + i\chi''$, is derived as

$$F = \frac{\chi'}{4}\nabla I + \frac{\chi''}{2} I \nabla \varphi, \tag{8}$$

where $I = \mathbf{E}.\mathbf{E}^*$, $\mathbf{E}$ is the laser electric field amplitude and $\varphi = kz$, $k$ is the laser wavenumber. Knowing definition of polarizability as $\mathbf{p} = \chi \mathbf{E}$, $\mathbf{p}$ is the electric dipole moment of a NP, from Eq. (7), one can derive the following

$$\chi = \frac{3}{4\pi} V_p \left( \frac{-1 + \varepsilon_n / \varepsilon_h}{2 + \varepsilon_n / \varepsilon_h} \right), \tag{9}$$

where $V_p$ is the volume of a spherical NP and as we will see in section VI, permittivity of each individual NP can be obtained from the phenomenological classical theory of Lorentz-Durud as following

$$\varepsilon_n = \frac{k^2 c^2}{\omega^2} = \varepsilon_{int} + 1 - \frac{\omega_p^2}{\omega^2 + i\omega\Gamma - \omega_p^2/3}, \tag{10}$$

where first term $\varepsilon_{int}$ is related to the contribution of bounded electrons (interband transitions) to the dielectric constant obtained by Lorentz-Durud theory, i.e. second two terms of the right hand side of the dielectric constant, in which only the dynamics of conduction electrons are taken into account, $\omega_p = (4\pi n_0 e^2 / m)^{1/2}$ is the electron plasma frequency, $\Gamma$ is the damping factor related to electron scattering, $e$ is the magnitude of electron charge and $m$ is the electron mass. In order to not to confuse readers, procedure of determining $\varepsilon_{int}$ and $\Gamma$ as a function of frequency will be explained in section IV. Optical force of Eq. (8) can make NPs to move and for their motion, we can use following continuity equation

$$\frac{\partial \rho_m}{\partial t} + \nabla \cdot \mathbf{J}_m = 0, \tag{11}$$

where $\rho_m$ is the mass density of NPs and $\mathbf{J}_m$ is their mass current density that is described by the Nernst-Planck equation [34]

$$\mathbf{J}_m = \rho_m \mathbf{v}_c - D \nabla \rho_m, \tag{12}$$

where $D$ represents diffusion coefficient and $\mathbf{v}_c$ is convective velocity of NPs which is related to the external force F acting on NPs through the relation $\mathbf{v}_c = \mu \mathbf{F}$ where $\mu$ is the mobility of NP. The first term of the right hand side of Eq. (12) is the drift current caused by the external force while the second one is the diffusion current resulted from Brownian motion. By combining Eqs. (11) and (12), we can obtain so-called Smoluchowski equation as the following

$$\frac{\partial \rho_m}{\partial t} + \nabla \cdot (\rho_m \mathbf{v}_c - D \nabla \rho_m) = 0. \tag{13}$$

Under steady state regime in which $\frac{\partial \rho_m}{\partial t} = 0$ and additionally in the equilibrium condition when drift is balanced by diffusion, the current density of particles is zero and we can write

$$\rho_m \mathbf{v}_c = D \nabla \rho_m, \tag{14}$$

Using the relation $\mathbf{v}_c = \mu \mathbf{F}$ and Eq. (8) for the external force, Eq. (14) reduces to the following

$$\frac{\chi \mu}{4} \rho_m \nabla I = D \nabla \rho_m, \tag{15}$$

which after integrating and using Einstein's relation, $D/\mu = k_B T$, $k_B$ is Boltzman constant and $T$ is the temperature, Eq. (15) leads to

$$\rho_m = \rho_0 \exp\left(\frac{\chi' I}{4k_B T} + \frac{\chi'' I}{2k_B T}\varphi\right), \tag{16}$$

where $\rho_0$ is the initial unperturbed mass density of NPs. It is worth mentioning that Eq. (16) can be used for number density of NPs, only by changing $\rho_0$ into $n_{c0} = 1/d^3$.

### II-C. Macroscopic electron current density

For a small non-relativistic NP subjected to low-intensity EM fields and whose radius is much less than wavelength $a \ll \lambda$, spatial variation of EM fields inside the particle is negligible as a result of which we can suppose that all electrons experience the same forces at a moment and whole electronic cloud moves together without any change in the shape and consequently the motion of an electron can present the motion of all electrons in the electronic cloud. Therefore, the equation describing the interaction of the EM fields of the laser with the electron cloud of each metallic nanoparticle can be written in the non-relativistic approximation as

$$\frac{d}{dt}(\mathbf{v}) + \frac{\omega_p^2}{3}\mathbf{r} + \Gamma\mathbf{v} = -\frac{e}{m}\left\{\mathbf{E}(\mathbf{r_0}) + (\mathbf{r}.\nabla)\mathbf{E}(\mathbf{r_0}) + \frac{1}{c}\mathbf{v}\times[\mathbf{B}(\mathbf{r_0}) + (\mathbf{r}.\nabla)\mathbf{B}(\mathbf{r_0})]\right\}, \tag{17}$$

where $\mathbf{v}$ is the velocity, $\mathbf{r}$ is the displacement of the electron cloud from the equilibrium state, additionally, the terms $\mathbf{E}(\mathbf{r_0})$ and $\mathbf{B}(\mathbf{r_0})$ represent the microscopic EM fields in the center of the particle at the position $\mathbf{r_0}$. In this equation, the term $-\mathbf{r}\omega_p^2/3$ relates to the restoration force induced by the displacement of the electron cloud with respect to the ions. Now we use the well-known perturbative method to solve the equation of motion of the electron cloud.

The first-order momentum equation can be easily obtained from Eq. (16) as

$$\frac{dv_x^{(1)}}{dt} + \frac{\omega_p^2}{3}x^{(1)} + \Gamma v_x^{(1)} = -\frac{\alpha e \hat{E}}{2m}e^{-i\omega t + ikz} + c.c., \tag{18}$$

where the superscript (1) refers to the first-order perturbation parameter and also coefficient appears because of the variation of laser's fields inside the metallic nanoparticle. It is worth mentioning that we used microscopic electric field $\alpha\mathbf{E}_L$ inside nanoparticle instead of the laser macroscopic field $\mathbf{E}_L$ and it may cause confusion because equations used for deriving amplifying/damping coefficient $\alpha$ were on the basis of the electrostatic approach that here cannot be valid for an electrodynamic problem. However, for a small particle whose radius is much less than wavelength $a \ll \lambda$, its representation as an ideal dipole is valid in the quasi-static regime, i.e. allowing for time-varying fields but neglecting spatial retardation effects over the particle volume [35, 36]. The solution of Eq. (17) is given by

$$x^{(1)} = \frac{1}{2} \frac{\alpha \hat{a} c \omega}{\omega^2 + i\omega\Gamma - \frac{1}{3}\omega_p^2} e^{-i\omega t + ikz} + c.c., \quad z^{(1)} = y^{(1)} = 0, \tag{19}$$

where $\hat{a} = e\hat{E}/mc\omega$ is the normalized laser amplitude.

The second-order equation can be written as

$$\frac{d\mathbf{v}^{(2)}}{dt} + \frac{\omega_p^2}{3}\mathbf{r}^{(2)} + \Gamma\mathbf{v}^{(2)} = -\frac{e}{mc}\mathbf{v}^{(1)} \times \mathbf{B}^{(1)}(\mathbf{r_0}). \tag{20}$$

Using Eqs. (2) and (18) we find

$$\mathbf{v}^{(1)} \times \mathbf{B}^{(1)}(\mathbf{r_0}) = \hat{\mathbf{z}} \frac{\alpha^2 \hat{a} c^2 k \omega \hat{E}}{4i(\omega^2 + i\omega\Gamma - \omega_p^2/3)} e^{-2i\omega t + 2ikz} + c.c. \tag{21}$$

Using Eq. (20) in Eq. (19), we find that

$$z^{(2)} = \frac{\alpha^2 \hat{a}^2 c^2 k \omega^2}{4i(\omega^2 + i\omega\Gamma - \omega_p^2/3)(4\omega^2 + 2i\omega\Gamma - \omega_p^2/3)} e^{-2i\omega t + 2ikz} + c.c. \tag{22}$$

We see that the second-order velocity of the electron clouds and the laser propagation direction are parallel. Macroscopically, such a velocity can produce longitudinal modulation of the macroscopic electron density. To derive the second-order perturbation of the electron density, we use the continuity equation

$$\frac{\partial n^{(2)}}{\partial t} + n_0 \nabla \cdot \mathbf{v}^{(2)} = 0. \tag{23}$$

Actually, to obtain the macroscopic electron density and current density, we should multiply both sides of the continuity equation (23) by the factor $4\pi l/3$, where $l = (a/d)^3$. Considering a solution in the form $n^{(2)} = (1/2)\tilde{n}e^{-2i\omega t+2ikz} + c.c.$, where $\tilde{n}$ is a complex amplitude, from the continuity equation we derive

$$n^{(2)} = -\frac{\alpha^2 \hat{a}^2 k^2 c^2 \omega^2}{2(\omega^2 + i\omega\Gamma - \omega_p^2/3)(4\omega^2 + 2i\omega\Gamma - \omega_p^2/3)} n_0 e^{-2i\omega t+2ikz} + c.c.\ . \qquad (24)$$

Such a density modulation leads to the creation of a longitudinal electric field of the second harmonic that can be obtained using Poisson's equation, $\nabla \cdot \mathbf{E}^{(2)} = -4\pi e n^{(2)}$, in the form

$$E_z^{(2)} = \frac{\alpha^2 \pi n_0 e \hat{a}^2 k c^2 \omega^2}{i(\omega^2 + i\omega\Gamma - \omega_p^2/3)(4\omega^2 + 2i\omega\Gamma - \omega_p^2/3)} e^{-2i\omega t+2ikz} + c.c.\ . \qquad (25)$$

The third-order equation of the electron cloud movement can be obtained as

$$\frac{d}{dt}\mathbf{v}^{(3)} + \frac{\omega_p^2}{3}\mathbf{r}^{(3)} + \Gamma\mathbf{v}^{(3)} = -\frac{e}{m}\left[\frac{\mathbf{v}^{(2)} \times \mathbf{B}^{(1)}(\mathbf{r_0})}{c} + (\mathbf{r}^{(2)} \cdot \nabla)\mathbf{E}^{(1)}(\mathbf{r_0})\right], \qquad (26)$$

where

$$-\frac{e}{m}(\mathbf{r}^{(2)} \cdot \nabla)\mathbf{E}^{(1)}(\mathbf{r_0}) = \frac{-\alpha^3 \hat{a}^3 c^3 \omega^3 k^2 \hat{\mathbf{x}}}{8(\omega^2 + i\omega\Gamma - \frac{1}{3}\omega_p^2)(4\omega^2 + 2i\omega\Gamma - \frac{1}{3}\omega_p^2)}\left(e^{-3i\omega t+3ikz} - e^{-i\omega t+ikz}\right) + c.c.\ . \qquad (27)$$

Using the lower-order parameters (Eqs. (22) and (27)) in Eq. (26), we can obtain the result

$$x^{(3)} = \frac{3\alpha^3 \hat{a}^3 c\omega^3 e^{-3i\omega t+3ikz}}{8(\omega^2 + i\omega\Gamma - \frac{\omega_p^2}{3})(9\omega^2 + 3i\omega\Gamma - \frac{\omega_p^2}{3})}\left(\frac{k^2 c^2}{(4\omega^2 + 2i\omega\Gamma - \frac{\omega_p^2}{3})}\right)$$
$$+ \frac{\hat{a}^3 \alpha^3 c\omega^3 e^{-i\omega t+ikz}}{8(\omega^2 + i\omega\Gamma - \frac{\omega_p^2}{3})^2}\left(\frac{k^2 c^2}{(4\omega^2 + 2i\omega\Gamma - \frac{\omega_p^2}{3})}\right) + c.c. \qquad (28)$$

Now, using the different orders of velocity and density for conduction electrons of individual NPs and own distribution of NPs from Eq. (16), finally, macroscopic electron current density will be obtained as

$$\mathbf{J} = -eV_p\left(n_0 + n^{(2)}\right)\left(\mathbf{v}^{(1)} + \mathbf{v}^{(3)}\right)n_{c0}\exp\left(\frac{\chi' I}{4k_B T} + \frac{\chi'' I}{2k_B T}\varphi\right). \tag{29}$$

**II-D. Nonlinear wave equation**

Some straightforward mathematical operations on Maxwell equations leads to the following nonlinear wave equation for the evolution laser pulse in a suspension of NPs

$$(\nabla^2 - \frac{\varepsilon_{eff}}{c^2}\frac{\partial^2}{\partial t^2})\mathbf{E} = \frac{4\pi}{c^2}\frac{\partial \mathbf{J}}{\partial t}, \tag{30}$$

where in the right hand side of the wave equation we should use current density of Eq. (29) and $\varepsilon_{eff}$ is the effective permittivity in which permittivity of ambient liquid and permittivity of only bound electrons of NPs are included in the calculation of effective permittivity by Maxwell Garnett [37] theory that with details will be explained later. Writing wave equation (30) only for first harmonic, after some uncomplicated algebraic operations, one can obtain

$$\left(\nabla^2 + \frac{\omega^2}{c^2}\varepsilon\right)\mathbf{E} = 0, \tag{31}$$

where

$$\varepsilon = \varepsilon_0 + \Phi(EE^*) - i\varepsilon_i, \tag{32}$$

$$\varepsilon_0 = Re(\varepsilon_{eff}) - \omega_P^2 n_{c0} V_p Re\left(\frac{1}{\omega^2 + i\omega\Gamma - \omega_P^2/3}\right), \tag{33}$$

$$-\varepsilon_i = Im\,\varepsilon_{eff} - \omega_P^2 n_{c0} V_p Im\left(\frac{1}{\omega^2 + i\omega\Gamma - \omega_P^2/3}\right), \tag{34}$$

and

$$\Phi = c_1|\hat{a}|^2 + c_2|\hat{a}|^4, \tag{35}$$

where we have

$$c_1 = -\frac{\omega^2\alpha^2\omega_P^2 n_{c0}V_p}{4(\omega^2+i\omega\Gamma-\omega_P^2/3)^2}\left(\frac{k^2c^2}{4\omega^2+2i\omega\Gamma-\omega_P^2/3}\right) - \frac{\omega_P^2 n_{c0}V_p}{\omega^2+i\omega\Gamma-\omega_P^2/3}\left(\frac{\chi'}{4k_BT}+\frac{\chi''}{2k_BT}\varphi\right)\left(\frac{mc\omega}{e}\right),$$

$$c_2 = -\frac{\omega^2\alpha^2\omega_P^2 n_{c0}V_p}{4(\omega^2+i\omega\Gamma-\omega_P^2/3)^2}\left(\frac{k^2c^2}{4\omega^2+2i\omega\Gamma-\omega_P^2/3}\right)\left(\frac{\chi'}{4k_BT}+\frac{\chi''}{2k_BT}\varphi\right)\left(\frac{mc\omega}{e}\right).$$

For deriving Eq. (35) we assumed that the norm of argument of exponential function, i.e. $\frac{\chi'I}{4k_BT}+\frac{\chi''I}{2k_BT}\varphi$, is much less than unity so we can expand it around zero and keep only squared terms of $\hat{a}$.

### II-E. Effective permittivity of medium

Now we calculate the effective permittivity of a background liquid medium with permittivity of $\varepsilon_h$ which includes metallic NPs. We use Maxwell Garnett method [37] to calculate the effective permittivity. For this purpose, first we determine the permittivity of an individual metallic NP. Contribution of conduction electrons on the NPs permittivity has been already considered in the nonlinear current density, and therefore we will not take into account these electrons calculating NP's permittivity. From Durud's model, considering only relatively free conduction electrons of a metallic bulk medium, permittivity can be written as

$$\varepsilon^f = 1 - \frac{\omega_p^2}{(\omega^2+i\Gamma\omega)}. \tag{36}$$

Considering contribution of other bound electrons, term $\varepsilon^b$ should be added to Eq. (36). Also, because of appearing restoration force in the case of limited dimension of NP, term $-\omega_p^2/3$ must be considered in the denominator. Therefore, we can write following expression for the permittivity of an individual NP

$$\varepsilon_n = \varepsilon_{int} + 1 - \frac{\omega_p^2}{\omega^2+i\Gamma\omega-\frac{\omega_p^2}{3}}. \tag{37}$$

Later, in the numerical discussions, we will use existing experimental [38] and theoretical data [39] for determining $\varepsilon_{int}$, $\omega_p$, and $\Gamma$ of silver NP. Now Maxwell Garnett approach can be used for calculating effective dielectric coefficient of metallic nano-suspension. For a host medium with dielectric constant $\varepsilon_h$ that includes metallic nanoparticles with dielectric constants $\varepsilon_n$ effective dielectric constant of medium, $\varepsilon_{eff}$, can be extracted from the following equation [37]

$$\frac{\varepsilon_{eff} - \varepsilon_h}{\varepsilon_{eff} + 2\varepsilon_h} = f \frac{\varepsilon_n - \varepsilon_h}{\varepsilon_n + 2\varepsilon_h}, \tag{38}$$

where $f = 4\pi l/3$ is the filling factor (the volume fraction occupied by a NP), $l = (r_c/d)^3$, $r_c$ is the average radius of NPs and $d$ is the average separation of NPs. As it has been mentioned earlier, we consider only the bound electrons contribution on the permittivity and set $\varepsilon_n = \varepsilon_{int}$.

Considering scattering of electrons in the motion equation leads to the appearance of imaginary part in the permittivity of an individual NP which in turn leads to EMW absorption in particles. Laser power absorbed by a nanoparticle can be written as $P_{abs} = \sigma_{abs} I_0$ where $\sigma_{abs}$ is the absorption cross section of particle and $I_0$ is the laser intensity. Using Mie theory [40] for particle whose size is much less than wavelength, absorption cross section is

$$\sigma_{abs} = 12\pi r_c^3 \frac{\omega}{c} \frac{\varepsilon''}{(\varepsilon' + 2)^2 + \varepsilon''^2}, \tag{39}$$

where $\varepsilon_n = \varepsilon' + i\varepsilon''$ is the permittivity of graphite nanoparticle with specific orientation whose real and imaginary part can be derived from Eq. (37) as following

$$\varepsilon' = 1 + \varepsilon_{int} - \frac{4\pi l}{3} \frac{\omega_p^2 (\omega^2 - \omega_p^2/3)}{(\omega^2 - \omega_p^2/3)^2 + \omega^2 \Gamma^2}, \tag{40}$$

$$\varepsilon'' = \frac{4\pi l}{3} \frac{\omega_p^2 \omega \Gamma}{(\omega^2 - \omega_p^2/3)^2 + \omega^2 \Gamma^2}. \tag{41}$$

**III. Envelope evolution**

Using oscillational form of Eq. (1) for the electric field of laser and substituting it in the wave equation (31), one can obtain following equation for nonlinear dynamics of laser amplitude

$$\left(\nabla_\perp^2 \hat{a} + 2ik\frac{\partial \hat{a}}{\partial z} + \frac{\partial^2 \hat{a}}{\partial z^2} - k^2 \hat{a}\right) + \frac{\omega^2}{c^2}\left(\varepsilon_0 - i\varepsilon_i + \Phi(\hat{a}\hat{a}^*)\right)\hat{a} = 0. \tag{42}$$

Now we exert the slow-variation condition of amplitude and omit the term $\frac{\partial^2 \hat{a}}{\partial z^2}$, also assume that $k$ is a real wave number which satisfies the following dispersion equation

$$k^2 = \frac{\omega^2}{c^2}\varepsilon_0 = \frac{\omega^2}{c^2}\left[Re(\varepsilon_{eff}) - \omega_P^2 n_{c0} V_p Re\left(\frac{1}{\omega^2 + i\omega\Gamma - \omega_P^2/3}\right)\right], \tag{43}$$

and simplify Eq. (42) as following

$$\left(\nabla_\perp^2 \hat{a} + 2ik\frac{\partial \hat{a}}{\partial z}\right) + \frac{\omega^2}{c^2}\left(-i\varepsilon_i + \Phi(\hat{a}\hat{a}^*)\right)\hat{a} = 0. \tag{44}$$

To study the problem of SF, we use the same procedure of [41]. First, we introduce dimensionless complex amplitude as following

$$\hat{a} = a_0(r,z) e^{-ikS(r,z)}. \tag{45}$$

Substituting Eq. (45) in Eq. (44) and separating its real and imaginary part results in two following coupled differential equations for real parameters of amplitude $a_0$ and beam eikonal $S$

$$\frac{1}{k^2 a_0}\nabla_\perp^2 a_0 - \left(\frac{\partial S}{\partial r}\right)^2 + 2\frac{\partial S}{\partial z} + \frac{1}{\varepsilon_0}Re\left[\Phi(a_0^2)\right] = 0, \tag{46}$$

$$a_0^2 \nabla_\perp^2 S + \left(\frac{\partial a_0^2}{\partial r}\right)\left(\frac{\partial S}{\partial r}\right) - \frac{\partial a_0^2}{\partial z} + \varepsilon_i \frac{k}{\varepsilon_0} a_0^2 - \frac{k}{\varepsilon_0} Im\left[\Phi(a_0^2)\right] a_0^2 = 0. \tag{47}$$

By inspiring from [42] we introduce following solutions for amplitude and beam eikonal in non-paraxial ray regime

$$a^2 = \frac{E_0^2}{f^2}\left(1 + a_2 \frac{r^2}{f^2 r_0^2} + a_4 \frac{r^4}{f^4 r_0^4}\right)\exp(-\frac{r^2}{f^2 r_0^2})e^{-2k_i z}, \tag{48}$$

$$S = S_0(z) + S_2(z)\frac{r^2}{r_0^2} + S_4(z)\frac{r^4}{r_0^4}. \tag{49}$$

By using Eqs. (48) and (49) in Eq. (35), one can obtain

$$\Phi = \Phi_0 + \Phi_2 + \Phi_4, \tag{50}$$

where for simplicity, we used the following definitions

$$\Phi_0 = c_1 \frac{E_0^2}{f^2} e^{-2kiz} + c_2 \frac{E_0^4}{f^4} e^{-4kiz}, \tag{51}$$

$$\Phi_2 = c_1 \frac{E_0^2}{f^2}(a_2 - 1)\frac{r^2}{f^2 r_0^2} e^{-2kiz} + c_2 \frac{E_0^4}{f^4}(2(a_2 - 1))\frac{r^2}{f^2 r_0^2} e^{-4kiz}, \tag{52}$$

$$\Phi_4 = c_1 \frac{E_0^2}{f^2}\left(-a_2 + a_4 + \frac{1}{2}\right)\frac{r^4}{f^4 r_0^4} e^{-2kiz} + c_2 \frac{E_0^4}{f^4}\left(2\left(a_4 - a_2 + \frac{1}{2}\right) + (a_2 - 1)^2\right)\frac{r^4}{f^4 r_0^4} e^{-4kiz}. \tag{53}$$

Replacing Eqs (48)-(50) in Eqs. (46) and (47), rearranging them in successive powers of $r^2$ and equating their coefficient to zero lead to the following equations for the unknowns of problem

$$\frac{ds_0}{dz} = \frac{1}{k^2 f^2 r_0^2}(1 - a_2) - \frac{1}{2\varepsilon_0} Re\Phi_0, \tag{54}$$

$$\frac{ds_2}{dz} = -\frac{1}{2k^2 f^4 r_0^2}(-3a_2^2 + 8a_4 - 2a_2 + 1) + \frac{2}{r_0^2} s_2^2 - \frac{r_0^2}{2\varepsilon_0 r^2} Re\Phi_2, \tag{55}$$

$$\frac{ds_4}{dz} = -\frac{1}{2k^2 f^2 r_0^2}(-14a_2 a_4 + 4a_2^3 + 2a_2^2 - 4a_4) + \frac{8}{r_0^2} s_2 s_4 - \frac{r_0^4}{2\varepsilon_0 r^4} Re\Phi_4, \tag{56}$$

$$\frac{df}{dz} = -\frac{f}{2}\left(\frac{4s_2}{r_0^2}\right) - \frac{k/\varepsilon_0}{E_0^2/f^2} Im\Psi_0, \tag{57}$$

$$\frac{da_2}{dz} = \frac{16f^2}{r_0^2} s_4 + \frac{4s_2}{r_0^2}(2a_2 - 1) + \frac{1}{f}\frac{df}{dz}(4a_2 - 2) - \frac{k/\varepsilon_0}{E_0^2/f^2} \text{Im}\left(\frac{f^2 r_0^2}{r^2}\Psi_2 + \Psi_0\right), \tag{58}$$

$$\frac{da_4}{dz} = \frac{1}{r_0^2}(12s_2 a_4 - 4s_2 a_2) + \left((6a_4 - 2a_2)\frac{1}{f}\frac{df}{dz}\right) + \frac{f^2}{r_0^2}(24s_4 a_2 - 4s_4)$$
$$- \frac{k/\varepsilon_0}{E_0^2/f^2}\text{Im}\left(\frac{f^4 r_0^4}{r^4}\Psi_4 + \frac{f^2 r_0^2}{r^2}\Psi_2 + \frac{1}{2}\Psi_0\right) \tag{59}$$

where we used following definitions

$$\Psi_0 = \left(c_1 + c_2 \frac{E_0^2}{f^2}e^{-2kiz}\right)\frac{E_0^4}{f^4}e^{-2kiz}, \tag{60}$$

$$\Psi_2 = \left[2c_1(a_2 - 1) + 3c_2 \frac{E_0^2}{f^2}(a_2 - 1)e^{-2kiz}\right]\frac{E_0^4}{f^4}\frac{r^2}{f^2 r_0^2}e^{-2kiz}, \tag{61}$$

$$\Psi_4 = \left[c_1\left((a_2 - 1)^2 + 2\left(a_4 - a_2 + \frac{1}{2}\right)\right) + c_2 \frac{E_0^2}{f^2}\left(3(a_2 - 1)^2 + 3\left(a_4 - a_2 + \frac{1}{2}\right)\right)e^{-2kiz}\right]\frac{E_0^4 r^4}{f^8 r_0^4}e^{-2kiz}, \tag{62}$$

Initial conditions for solving Eqs. (54)-(59) are as following

$$f(z=0) = 1, \; S_i(z=0) = 0, \; a_j(z=0) = 0, \; i = 0,2,4 \; \& \; j = 2,4. \tag{63}$$

**IV. Numerical discussions**

A numerical analysis has been carried out to solve equations (54)-(55) with initial conditions of Eq. (63) in order to find spatial evolutions of laser and distribution of NPs of suspension as well. We consider silver NPs with $n_0 = 5.86 \times 10^{22} cm^{-3}$ and radius $r_c = 15nm$. To make an evaluation for parameter $l$, let us consider $d = 2r_c, 3r_c, 4r_c$ that leads to the following values of $l = 0.125$, 0.037 and 0.016, respectively. The first case ($l = 0.125$) is related to the kissing contact of two NPs and it cannot be realistic. For all cases, we set $T = 293\,^0K$, $r_0 = 15\mu m$, $l = 0.016$ and longitudinal distance z has been normalized by Rayleigh length $Z_R = k r_0^2/2$.

Figure 1 presents numerical solution of linear dispersion for the suspension of gold NPs in water. On the vertical axis, $\bar{k}_r$ and $\bar{k}_i$ represent the real and imaginary part of the wave number normalized by $\omega_p/c$, respectively and horizontal axis is the normalized frequency $\omega/\omega_p$. For the frequency area $\omega < 0.4\omega_p \approx 9.94 \times 10^{14} s^{-1}$, the imaginary part of the wavenumber is much less than the real part and the absorption is negligible. By the increase in frequency, $\bar{k}_i$ increases and becomes comparable with $\bar{k}_r$. It takes a maximum around the plasmon resonance of gold NP, i.e. $\omega/\omega_p \approx 1/\sqrt{3}$. In order to determine the frequency area in which laser focuses, we have plotted the variations of spot size parameter $f$ with respect to the normalized frequency in a fixed propagation distance $z/Z_R = .01$. For $\omega/\omega_p < 0.45$, $f < 1$ and laser focuses. The minimum value of $f$ that shows the best condition for the laser focusing, appears at frequency $\omega/\omega_p = 0.4$. Figure 3 shows the variations of $f$ with respect to the normalized propagation distance $z/Z_R$ when the laser initial maximum intensity at the beam center is $I_L = 10^{15} Wcm^{-2}$ (or $\hat{a}(z=0) = e\hat{E}/mc\omega = 3 \times 10^{-7}$) and $\omega = 0.4\omega_p \approx 9.94 \times 10^{14} s^{-1}$ (approximate equivalent wavelength in the vacuum is $\lambda \approx 0.4 \mu m$). In this figure, the spot size decreases with respect to the propagation length, reaches a minimum value of $f \approx 0.25$ at $z/Z_R \approx 0.018$ then starts increasing. Decrease of the spot size is equivalent to the beam focusing and its increase means defocusing of laser beam. Notice that laser is simultaneously attenuated and in order to be able speak about the wave propagation, wavelength should be much greater than the penetration depth, i.e. $\delta = 2\pi/k_i$, where $k_i$ is the imaginary part of the wave number. In this case, calculations show that $\delta \approx 90 \mu m$ and $Z_R \approx 3.5 mm$ then we have $\lambda, Z_R \gg \delta$. Figure 4 illustrates distribution of the normalized laser intensity in the medium. Laser intensity decreases with respect to the propagation length more rapidly that it was expected by the linear theory according to skin depth $\delta/2Z_R \approx 0.016$, on the symmetry axis and in $(z/Z_R) \approx 0.002$, intensity decreases to 0.4 of its maximum value that is much less than skin depth. It is because of the nonlinear force caused by the gradient of medium polarization, i.e. Eq. 8, which makes NPs to accumulate around the symmetry axis and it leads to the more attenuation of laser intensity. Figure 5 shows concentration of particles around z axis by the nonlinear force. In a fixed r, increase in z causes increase in the particles concentration then it reaches a maximum at $(z/Z_R) \approx 0.002$ and decreases. For all constant z increase in r leads to the decrease in the concentration of NPs. On the z axis on which we have the maximum particle distribution, at $(z/Z_R) \approx 0.002$, particles concentration becomes 2.5 times more than its equilibrium value. Displacement of particles distribution to nonzero value of z is the result of existing of imaginary part of particle's polarizability and wavenumber and of being function of z for perpendicular structure of laser field by Eq. (48) as well that cause occurrence of nonlinear force along the propagation direction. Longitudinal forces regarding to the first and second term of Eq. (8) are denoted by $F_{1z}$ and $F_{2z}$, respectively and are plotted with respect to the variations of coordinates z and r in Figs. 6-a and 6-b. Longitudinal force related to the existing of laser profile dependent on z, $F_{1z}$, is negative and it derives particles contemporary with respect to the pulse propagation however the second term of longitudinal force is positive and its absolute value is greater, therefore total longitudinal force derives particles in the same direction of laser pulse propagation and localizes them out of z=0 in the positive direction. Spatial variation of radial nonlinear force of first and second term of Eq. (8) is plotted in Figs. 7. It is clear that both terms of radial force are negative and they make particle localize around symmetry axis.

## V. Conclusions

We investigated SF of an intense laser pulse in a magnetized bulk medium of graphite NPs. Spot size evolution of right-handed circularly-polarized laser is studied. Effect of laser frequency, average separation of NPs, laser intensity and external magnetic field on the SF are considered. It is found that near to plasmon resonance frequency (at IR area around wavelength of several microns) a resonant focusing happens. In addition, it is observed that the increase in the ratio of NPs radius to their separation leads to the more focusing of laser. Furthermore, increase in the magnetic field causes improvement of focusing property of medium. It is numerically observed that threshold power of SF and effective magnetic field for increasing nonlinearity of graphite NPs medium are much lower than similar case of plasma medium. Therefore, the feasibility of using such nonlinear medium in experiments can have a good perspective in future.

**References:**

[1] S. A. Maier, P. G.. Kik, H. A. Atwater, H. Meltzer, E. Harel, B. E. Koel and A. A. G. Requicha, Nature Materials **2**, 229 (2003).

[2] M. L. Brongersma, J. W. Hartman, and H. A. Atwater, Phys. Rev. B **62**, R16356(R) (2000).

[3] D. K. Gramotnev and S. I. Bozhevolnyi, Nature Photonics **4**, 83(2010).

[4] J. R. Krenn, Nature Materials **2**, 210 (2003).

[5] P. K. Jain, K. S. Lee, I. H. El-Sayed, and M. A. El-Sayed, J. Phys. Chem. B **110**, 7238 (2006).

[6] D. Cotter, R. J. Manning, K. J. Blow, A. D. Ellis, A. E. Kelly, D. Nesset, I. D. Phillips, A. J. Poustie, D. C. Rogers, Science 19, 1523 (1999).

[7] S. Porel, N. Venkatram, D. N. Rao, T. P. Radhakrishnana, Optical power limiting in the femtosecond regime by silver nanoparticle-embedded polymer film. J. App. Phys. **102**, 033107 (2007).

[8] Yang L, Becker K, Smith F M, Marguder R H, Haglund R F, Yang L, Dorsinville R, Alfano R R and Zuhr R A 1994 J. Opt. Soc. Am. B 11 457–69

[9] K. Uchida , S. Kaneko, S. Omi, C. Hata, H. Tanji, Y. Asahara and A. J. Ikushima, J. Opt. Soc. Am. B **11**, 1236 (1994).

[10] F. Hache, D. Ricard and C. Flytzanis, J. Opt. Soc. Am. B **3**, 1647 (1986).

[11] S. C. Mehendale, S. R. Mishra , K. S. Bindra, M. Laghate, T. S. Dhami and K. S. Rustagi, Opt. Commun. **133**, 273 (1997).

[12] J. Parashar, Phys. Plasmas **16**, 093106 (2009).

[13] A. Chakhmachi and B. Maraghechi, Phys. Plasmas **18**, 022102 (2011).

[14] A. Chakhmachi, Phys. Plasmas **18**, 062104 (2013).

[15] N. Spehri Javan, J. App. Phys. **118**, 073104 (2015).

[16] A. Ahmad and V. K. Tripathi, Nanotechnology **18**, 315702 (2007).

[17] C. Min, P. Wang, X. Jiao, Y. Deng, H. Ming, Appl. Phys. B **90**, 97–99 (2008).

[18] H. A. Bethe, Phys. Rev. **66**, 163(1944).

[19] W. L. Barnes, W. A. Murray, J. Dintinger, E. Devaux, and T. W. Ebbesen, Phys. Rev. Lett. **92**, 107401(2004).

[20] J. Bravo-Abad, F. J. Garcia-Vidal, and L. Martin-Moreno, Phys. Rev. Lett. **93**, 227401 (2004).

[21] S.-H. Chang, S. K. Gray, and G. C. Schatz, Opt. Express **13**, 3150 (2005).

[22] F. J. Garcia-Vidal and L. Martin-Moreno, Phys. Rev. B **66**, 155412 (2002).

[23] J. R. Krenn, A. Dereux, J. C. Weeber, E. Bourillot, Y. Lacroute, and J. P. Goudonnet, G. Schider, W. Gotschy, A. Leitner, and F. R. Aussenegg, C. Girard, Phys. Rev. Lett. **89**, 2590 (1999).

[24] D. Y. Lei, A. Aubry, S. A. Maier and J. B. Pendry, New J. Phys. **12**, 093030 (2010).

[25] S. A. Akhmanov, A. P. Sukhurov, and R. V. Khokhlov, Sov. Phys. Usp. **10**, 609 (1968).

[26] N. Barsch, J. Jakobi, S. Weiler, and S. Barcikowski, Nanotechnology 20, 445603 (2009).

[27] R. A. Ganeev, A. I. Ryasnyansky, A. L. Stepanov, C. Marques, R. C. da Silva and E. Alves, Opt. Comm. **253**, 205 (2005).

[28] N. Sepehri Javan, Phys. Plasmas **22**, 093116 (2015).

[29] N. Sepehri Javan, F. Rouhi Erdi, M. N. Najafi, Phys. plasmas **24**, 052301 (2017).

[30] A. Ashkin, J.M. Dziedzic, J.E. Bjorkholm and S. Chu, "Observation of single-beam gradient force optical trap for dielectric particles," Opt. Lett. **11**, 288-290 (1986).

[31] El-Ganainy, R.; Christodoulides, D. N.; Rotschild, C.; Segev, M. Opt. Express 2007, **15**, 10207−10218.

[32] W. Man, S. Fardad, Z. Zhang, J. Prakash, M. Lau, P. Zhang, M. Heinrich, D. N. Christodoulides, and Z. Chen, Phys. Rev. Lett. **111**, 218302 (2013).

[33] L. Novotny, *Principles of Nano-Optics*, Cambridge University Press, Cambridge, UK (2006).

[34] B. J. Berne and R. Pecora, *Dynamic Light Scattering: With Applications to Chemistry, Biology and Physics*, Dover Publication, New York, USA (2000).

[35] S. A. Maier, *Plasmonics: Fundamentals and Applications*, Springer, New York, USA (2007).

[36] L. De Sio, *Active Plasmonic Nanomaterials*, Taylor & Francis Group, Florida, USA (2016).

[37] O. Levy and D. Stroud, Phys. Rev. B **56**, 8035(1997).

[38] E. D. Palik, *Handbook of Optical Constants of Solids*, Academic Press, NewYork, USA (1983).

[39] P. B. Johnson and R. W. Christy, Phys. Rev. B **6**, 4370 (1972).

[40] C. Bohren and D. Huffman, *Absorption and Scattering of Light by Small Particles. A Wiley-International Publication*, NewYork, USA (1983).

[41] S. T. Navare, M. V. Takale, S. D. Patil, V. J. Fulari, and M. B. Dongare, Opt. Lasers Eng. **50**, 1316 (2012).

[42] C. S. Liu and V. K. Tripathi, Phys. Plasmas **7**, 4360 (2000).

**Figure Captions**

Fig. 1 Variation of normalized laser spot size $r_s/r_0$ with respect to with respect to normalized propagation distance $z/Z_R$ for five different values of external magnetic field, when $I_L = 10^{15} Wcm^{-2}$, $\omega_0 = 9.94 \times 10^{14} s^{-1}$ and $l = (1/6)^3$.

Fig. 2 Variation of normalized laser spot size $r_s/r_0$ with respect to normalized propagation distance $z/Z_R$ for unmagnetized and magnetized ($B_0 = 100T$) cases, when $I_L = 10^{15} Wcm^{-2}$, $\omega_0 = 0.9 \omega_{p\perp}/\sqrt{3} \approx 7.49 \times 10^{14} s^{-1}$ and $l = (1/6)^3$.

Fig. 3 Variation of normalized laser spot size $r_s/r_0$ with respect to normalized propagation distance $z/Z_R$ for unmagnetized and magnetized ($B_0 = 100T$) cases, when $I_L = 10^{15} Wcm^{-2}$, $\omega_0 = \omega_{p\perp}/\sqrt{3} \approx 8.33 \times 10^{14} s^{-1}$ and $l = (1/6)^3$.

Fig. 4 Variation of normalized laser spot size $r_s/r_0$ with respect to normalized propagation distance $z/Z_R$ for unmagnetized and magnetized ($B_0 = 100T$) cases, when $I_L = 10^{15} Wcm^{-2}$, $\omega_0 = 1.1 \omega_{p\perp}/\sqrt{3} \approx 9.16 \times 10^{14} s^{-1}$ and $l = (1/6)^3$.

Fig. 5 Variation of normalized laser spot size $r_s/r_0$ with respect to normalized propagation distance $z/Z_R$ for three different values of average separation of nanoparticles, when $\omega_0 = 1.1 \omega_{p\perp}/\sqrt{3} \approx 9.16 \times 10^{14} s^{-1}$, $I_L = 10^{15} Wcm^{-2}$, $B_0 = 100T$.

Fig. 6 Variation of normalized laser spot size $r_s/r_0$ with respect to normalized propagation distance $z/Z_R$ for three different values of intensity, when $\omega_0 = 1.1 \omega_{p\perp}/\sqrt{3} \approx 9.16 \times 10^{14} s^{-1}$, $l = (1/5)^3$, $B_0 = 100T$.